\documentclass[prl,twocolumn,showpacs,floatfix]{revtex4}
\usepackage{graphicx}
\begin{document}

\title {Solitary waves in elongated clouds of strongly-interacting
bosons}
\author{M. \"{O}gren$^1$, G. M. Kavoulakis$^1$, and A. D. Jackson$^2$}
\affiliation{$^1$Mathematical Physics, Lund Institute of
Technology, P.O. Box 118, SE-22100 Lund, Sweden \\
$^2$Niels Bohr Institute, Blegdamsvej 17, DK-2100, Copenhagen
\O, Denmark}
\date{\today}

\begin{abstract}
We examine the propagation of solitary waves in elongated
clouds of trapped bosonic atoms as the confinement, the
strength of the interatomic interaction, and the atom density
are varied. We identify three different physical regimes and
develop a general formalism that allows us to interpolate
between them. Finally we pay special attention to the
transition to the Tonks-Girardeau limit of strongly-interacting
bosons.
\end{abstract}
 \maketitle

By appropriate manipulation of the trapping potential, confined
ultracold atoms can achieve conditions of reduced
dimensionality.  Taking advantage of this property, two recent
experiments described in Refs.\,\cite{Paredes,Science} have
managed to create conditions such that bosonic atoms approached
the so-called Tonks-Girardeau phase \cite{TGIR} predicted for
strongly-interacting bosons in one dimension.  In the
experiment of Ref.\,\cite{Science}, bosonic atoms were confined
in elongated traps, and the gas was observed to approach the
Tonks-Girardeau limit as the transverse confinement was
increased.  Evidence for this transition was obtained from
measurements of the energy and of the axial size of the gas. In
the experiment of Ref.\,\cite{Paredes} the momentum
distribution provided evidence for this transition.

Motivated by these experiments, we will consider here how
solitary waves emerge in strongly-interacting,
quasi-one-dimensional atomic gases.  We first present a general
description of the problem, identifying three physically
distinct regimes \cite{Liebnew}.  We then develop a formalism
that allows us to calculate the density profile, energy, and
momentum associated with grey/dark solitary waves
\cite{Lieb,Tsuzuki,JKP,Kol3,GW,KPa,JK,KPa2}.  This formalism
can be applied to any quasi-one dimensional system given only
knowledge of the energy per unit length as function of the
density per unit length.  We find interesting changes in the
properties of solitary waves as a function of density which
suggest that the study of solitary waves in these elongated
clouds could provide confirmation of the transition to the
Tonks-Girardeau limit \cite{Burger,Den}.  Our results also
suggest eventual technological applications, e.g., the
transmission of signals in atomic waveguides.

Here, we will neglect the effects of trapping along the
$z$-axis of weak confinement and will for simplicity consider a
cylindrical trap, $V = M \omega_{\perp}^2 (x^2 + y^2)/2$, where
$M$ is the atom mass, and $\omega_{\perp}$ is the frequency of
the trapping potential in the transverse direction.  We
approximate a short-ranged atom-atom interaction as $V_{\rm
int}({\bf r}-{\bf r}')= U_0 \delta({\bf r}-{\bf r}')$ where
$U_0 = 4 \pi \hbar^2 a/M$ and $a$ is the s-wave scattering
length for elastic atom collisions. The Gross-Pitaevskii
equation for the order parameter, $\Psi$, then has the form
\begin{eqnarray}
   i \hbar \, \partial_t \Psi =
     (-\hbar^2 {\bf \nabla}^2 / {2M} + U_0 |\Psi|^2 + V) \Psi.
\label{GPequ}
\end{eqnarray}
Following Ref.\,\cite{KP}, we assume that the transverse
dimension of the cloud is sufficiently small and the
corresponding time scale sufficiently short that the transverse
profile of the particle density can adjust to the equilibrium
form appropriate for the local atomic density.  With this
approximation, the problem becomes one-dimensional, and the
solitary pulse can be described by a local velocity, $v(z)$,
and a local density of particles per unit length,
$\sigma(z)=\int |\Psi(x,y,z)|^2 \, dx dy$ \cite{KP}.  The order
parameter can then be written as a simple product \cite{JKP}
with $\Psi({\bf r},t)=f(z,t) \, g(x,y,\sigma)$, where $g$ is
the equilibrium wavefunction for the transverse profile. If $g$
is chosen to be normalized, $\int |g|^2 dx dy = 1$, the
equations above imply that $|f|^2=\sigma$.

In this problem there are three qualitatively distinct physical
regimes and two corresponding transitions separating them.  The
first of these transitions involves a change in the transverse
profile of the cloud.  For weak transverse confinement, $|g|^2$
can be calculated using the Thomas-Fermi approximation and is
simply a parabola. For transverse confinement of a strength
sufficient to ensure that the chemical potential is much
smaller than $\hbar \omega_{\perp}$, $|g|^2$ becomes Gaussian.
The second transition involves a change in the longitudinal
profile of solitary waves.  For weak transverse confinement,
the atoms form a condensate.  In the limit of strong
interactions, strong transverse confinement and small linear
densities, however, bosonic atoms behave in a certain sense
like non-interacting fermions.  This transition involves no
change in the (Gaussian) transverse profile since the chemical
potential of the gas is much smaller than $\hbar
\omega_{\perp}$ in both cases.

We first consider the transition in the transverse direction.
The critical value of $\sigma a$ for this transition is
determined by the condition that the kinetic energy associated
with transverse motion, $\hbar^2/M R^2$, is comparable to the
typical interaction energy, $n U_0$, where $R$ is the
transverse width of the cloud and $n$ is the typical
(three-dimensional) density.  Since $n \sim N/(Z \pi R^2)$,
where $N/Z = \sigma$ is the density per unit length, one sees
that the value of $\sigma$ at the crossover is $\sigma_{c,1}
\sim 1/a$.  It is interesting to note that both energy scales
vary as $1/R^2$ with the consequence that $\sigma_{c,1}$ is
independent of the oscillator length $a_{\perp}=(\hbar/M
\omega_{\perp})^{1/2}$.

Under typical experimental conditions (i.e., where a single
trap rather than a series of tubes is used), $\sigma a \gg 1$.
The cloud is thus in the Thomas-Fermi regime.  The situation
changes, however, with greater transverse confinement. To see
this, we recall that the cloud expands along the long axis of
the trap when it is squeezed transversely.  Since $Z$
increases, $\sigma=N/Z$ decreases.

In the limit $\sigma a \ll 1$, $|g|^2$ has a Gaussian form,
$|g|^2 = {\rm e}^{-({\rho}/a_{\perp})^2}/(\pi a_{\perp}^2)$. As
shown in Ref.\,\cite{JKP}, $f$ then satisfies the equation
\begin{eqnarray}
   i \hbar \, \partial_t f =
     - (\hbar^2/2M) \partial_z^2 f +
      \hbar \omega_{\perp} (1+ 2 a |f|^2) f.
\label{fgp1w}
\end{eqnarray}
We see from this equation that $f \propto {\rm e}^{-i
\omega_{\perp} (1+2a \sigma_0) t}$ as $|z| \to \infty$, where
$\sigma_0$ is the background linear density.  Rewriting
Eq.\,(\ref{fgp1w}) using the variable $w=f e^{i \omega_{\perp}
(1+2a \sigma_0) t}$, we obtain
\begin{eqnarray}
   i \hbar \, \partial_t w =
     - (\hbar^2/2M) \partial_z^2 w +
      \hbar \omega_{\perp} 2 a (|w|^2 - \sigma_0) w.
\label{gp42w}
\end{eqnarray}
Equation (\ref{gp42w}) includes a familiar (i.e., quadratic)
nonlinear term and leads to a speed of sound, $c_{1}$, which
satisfies the equation $M c_{1}^2 = 2 \hbar \omega_{\perp}
\sigma_0 a$. Since $\sigma_0 = n_0 \pi a_{\perp}^2$, we see
that $M c_{1}^2 = n_0 U_0/2$ \cite{JKP}.

As shown in Refs.\,\cite{JKP,JK}, the density associated with a
solitary wave of velocity $u$ has the form
\begin{eqnarray}
   \sigma(z)/\sigma_0 - 1 = - {\cos^2 \theta}/
   {\cosh^2(z \cos \theta /\zeta)},
\label{solEull}
\end{eqnarray}
where $\theta = {\sin}^{-1}(u/c_{1})$ and $\zeta=2\xi(n_0)$.
Here $\xi(n_0)$ is the coherence length for $n_0 = \sigma_0/
(\pi a_{\perp}^2)$, so that $\zeta = a_{\perp}/ (2 \sigma_0
a)^{1/2}$.  Further, the dispersion relation connecting the
energy and momentum of a solitary wave, ${\cal E}({\cal P})$,
is given parametrically as ${\cal E}/{\cal E}_0 = (4 \sqrt 2/3)
\cos^3 \theta$, where ${\cal E}_0=\hbar \omega_{\perp}
(\sigma_0 a)^{1/2} \sigma_0 a_{\perp}$, and ${\cal P}/{\cal
P}_0 = \pi u / |u| - 2 \theta - \sin 2 \theta$, where ${\cal
P}_0 = \sigma_0 \hbar$.

We now consider the second, longitudinal transition.  As noted
above, the gas approaches the so-called Tonks-Girardeau limit,
in which the bosons to some extent behave like non-interacting
fermions, as the transverse confinement is increased.  The
motion of the atoms in this limit is effectively
one-dimensional since transverse degrees of freedom are frozen
out when $\hbar \omega_{\perp}$ is much larger than the
chemical potential.  In the crossover region, each atom
occupies a length of order $1/\sigma_0$ along the axis of the
trap.  The corresponding kinetic energy is on the order of
$\hbar^2 \sigma_0^2/M$. The typical interaction energy, on the
other hand, is on the order of $n_0 U_0$, where $n_0 \sim
\sigma_0/ a_{\perp}^2$ so that $n_0 U_0 \sim \hbar^2 \sigma_0
a/(M a_{\perp}^2)$. Thus, the ratio between the interaction
energy and the kinetic energy is $\sim a/(\sigma_0 a_{\perp}^2)
\sim [\sigma_0 \xi(n_0)]^{-2}$. Given the assumption of strong
interactions, this quantity is much larger than unity for small
values of $\sigma_0$ (i.e., for low densities), small values of
$a_{\perp}$ (i.e., for strong transverse confinement), or large
values of $a$ (i.e., for strong interactions).  The transition
between the two regimes takes place for $\sigma_{c,2} \sim
a/a_{\perp}^2$.

For values of $\sigma_0$ much smaller than $\sigma_{c,2}$
\cite{Kol2},
\begin{eqnarray}
   i \hbar \, \partial_t f =
         (\hbar^2/2M) (- \partial_z^2 f + \pi^2 |f|^4 f).
\label{ton1}
\end{eqnarray}
In this case we see that $f \propto e^{-i (\pi^2 \sigma_0^2
\hbar/2M) t}$ as $|z| \to \infty$.  We thus rewrite
Eq.\,(\ref{ton1}) using the variable $w=f {\rm e}^{i (\pi^2
\sigma_0^2 \hbar/2M) t}$ to obtain
\begin{eqnarray}
   i \hbar \, \partial_t w =
         (\hbar^2/2M) [- \partial_z^2 w +
         \pi^2 (|w|^4 - \sigma_0^2) w].
\label{ton2}
\end{eqnarray}
The speed of sound $c_{2}$ is now given as $M c_{2}^2 = \pi^2
\hbar^2 \sigma_0^2/M$ or $c_{2} = \pi \hbar \sigma_0 /M$.  This
is precisely the Fermi velocity in one dimension with $k_F =
\sigma_0 \pi$.

As shown in Ref.\,\cite{Kol3}, the density of the cloud
associated with the solitary wave is now
\begin{equation}
   \sigma(z)/\sigma_0 - 1 = - \frac {3 \cos^2 \theta}
      {2+(1 + 3 \sin^2 \theta)^{1/2}
      \cosh(2 \pi \sigma_0 z \cos \theta)},
\label{ssolEull}
\end{equation}
where $\theta = {\sin}^{-1}(u/c_{2})$.  The dispersion relation
can again be expressed parametrically using
\begin{eqnarray}
  {\cal E}/{\cal E}_0 =
    \frac {\sqrt 3 \pi} {2}
         \cos^2 \theta \, \ln \left[
         \frac {2 + 3 \cos^2 \theta} {(1 + 3 \sin^2 \theta)^{1/2}}
         \right],
\label{eew2fs}
\end{eqnarray}
with ${\cal E}_0=\hbar^2 \sigma_0^2/M$, and
\begin{equation}
  {\cal P} /
           {\cal P}_0 =   - \pi ({\cal E}/{\cal E}_0)
 \, \tan \theta + \cos^{-1} \left[ \frac {3 \sin^2 \theta - 1}
      {(1 + 3 \sin^2 \theta)^{1/2}} \right].
\label{pst1133344}
\end{equation}

It is natural to ask how one actually proceeds from one region
to another.  The transverse transition was studied in some
detail in Refs.\,\cite{JKP,KPa2}.  Here, we focus on the
second, longitudinal transition using a formalism of more
general applicability.  The basic ingredient required is the
equation of state, i.e., the energy of the gas per unit length,
$\epsilon(\sigma)$, as function of $\sigma$. This can be
calculated from the Lieb-Liniger model \cite{LL} and can be
written as $\epsilon(\sigma) = (\hbar^2 \sigma^3/ 2 M) \,
\tilde{e}(\gamma)$, where $\tilde{e}(\gamma)$ is a numerically
known function and $\gamma = 2 a / (\sigma a_{\perp}^2)$.

Given $\epsilon(\sigma)$, one can immediately determine the
sound velocity as $ M c^2 = \sigma_0 \partial^2
\epsilon/\partial \sigma^2(\sigma_0)$.  The energy of the
solitary wave is
\begin{eqnarray}
   {\cal E} = \int \left( \frac {\hbar^2} {2 M}
        \frac {\partial w^*} {\partial z} \frac {\partial w}
   {\partial z} + \epsilon(\sigma) - \sigma \epsilon'(\sigma_0) + C
   \right) dz
\label{ew1}
\end{eqnarray}
with $\epsilon' = \partial \epsilon/\partial \sigma$.  The
final term in this equation represents the energy of the
background density of atoms and ensures convergence of the
integral. Thus, $C = -\epsilon(\sigma_0) + \sigma_0
\epsilon'(\sigma_0)$.

Equation (\ref{ew1}) implies that $w$ satisfies the equation
\begin{eqnarray}
   i \hbar \, \partial_t w =
      - (\hbar^2/2M) \partial_z^2 w + [\epsilon'(\sigma) -
           \epsilon'(\sigma_0)] w.
\label{ton5}
\end{eqnarray}
Writing $w=\sqrt \sigma \, {\rm e}^{i \phi}$ and separating the
real and imaginary parts of Eq.\,(\ref{ton5}), we obtain the
two hydrodynamic equations
\begin{eqnarray}
   \frac {\hbar^2} {2M}
      \left( \frac {\partial \sqrt \sigma} {\partial z} \right)^2 =
      \epsilon(\sigma) - \epsilon(\sigma_0) -
                       (\sigma - \sigma_0) \epsilon'(\sigma_0)
        \nonumber \\
        - M u^2 \frac {(\sigma - \sigma_0)^2} {2\sigma}
\label{Eul1}
\end{eqnarray}
and $v = (\hbar/M) \partial \phi / \partial z = u (1 - \sigma_0
/ \sigma)$. Here, we have imposed the boundary condition that
$v \to 0$ as $\sigma \to \sigma_0$. Eq.\,(\ref{ew1}) can be
written as
\begin{eqnarray}
{\cal E} = \int \left[ \frac {\hbar^2} {2 M} \left( \frac
{\partial \sqrt \sigma} {\partial z} \right)^2 + \frac {\hbar^2
\sigma} {2 M} \left( \frac {\partial \phi} {\partial z} \right)^2
+ \epsilon(\sigma) - \epsilon(\sigma_0) \right. \nonumber \\
 - (\sigma - \sigma_0) \epsilon'(\sigma_0)] \, dz. \phantom{X}
\label{ew2}
\end{eqnarray}
Combining the above equations, we obtain
\begin{equation}
{\cal E} = 2 \int \left[ \epsilon(\sigma) - \epsilon(\sigma_0) -
(\sigma - \sigma_0) \epsilon'(\sigma_0) \right] dz.
\label{een}
\end{equation}
Finally, the momentum of the solitary wave is given as
\begin{equation}
 {\cal P} = M \int (\sigma - \sigma_0) v(z) \, dz
  = M u \int \frac {(\sigma-\sigma_0)^2} {\sigma} \, dz.
\label{pst2a}
\end{equation}

We thus see that, given knowledge of the energy per unit length
$\epsilon(\sigma)$, Eq.\,(\ref{Eul1}) allows us to determine
the shape of the solitary wave $\sigma(z)$ for a given
velocity, $u$.  Equations (\ref{een}) and (\ref{pst2a}) then
give ${\cal E}(u)$ and ${\cal P}(u)$, which can be combined to
establish the dispersion relation, ${\cal E} = {\cal E}({\cal
P})$. While the profiles of the solitary wave and the
corresponding dispersion relations are known in all three
limits examined earlier \cite{Lieb,Tsuzuki,Kol3,KPa,JK,KPa2},
Eqs.\,(\ref{ton5}) -- (\ref{pst2a}) allow us to interpolate
between them.  As indicated, our approach is quite general and
merely requires knowledge of $\epsilon(\sigma)$.

Let us now consider a specific example for the case $\gamma_0 =
2 a /(\sigma_0 a_{\perp}^2)= 1$.  It is convenient to express
$\epsilon$ explicitly in terms of $\sigma$ with the result that
$\epsilon(\sigma) = [\hbar^2 \sigma_0^3 / 2 M] \,
\tilde{e}(\sigma/\sigma_0)$, where a reliable interpolation
formula for $\tilde{e}$ is given as $\tilde{e}(y) = (\pi^2
\gamma_0^3 y^3/3 + \kappa \gamma_0 y^5)/(\gamma_0^3 + 4
\gamma_0^2 y + 4 \gamma_0 y^2 + \kappa y^3)$ with $y =
\sigma(z)/\sigma_0$ and $\kappa \approx 6.879$.

For $\gamma_0 = 1$, Eq.\,(\ref{Eul1}) then has the form
\begin{eqnarray}
   \left( \frac {\partial \sqrt y} {\partial \tilde z} \right)^2 =
     \frac {\pi^2 y^3/3 + \kappa y^5}  {1 + 4 y + 4 y^2 + \kappa y^3}
    - \frac {\pi^2 /3 + \kappa } {9 + \kappa}
    \nonumber \\
    - (y - 1) \,\, \partial \tilde{\epsilon}/ \partial y
    |_{y=1}
                - \left( \frac {M c}{\hbar \sigma_0} \right)^2
                \frac {u^2} {c^2}
         \frac {(y - 1)^2} y,
\label{EEul1}
\end{eqnarray}
where $\tilde z = \sigma_0 z$. Given that $\partial
\tilde{e}(y)/ \partial y \approx 1.471$ and $\partial^2
\tilde{e}(y) / \partial y^2 \approx 1.872$ for $y=1$, we find
that the sound velocity is given as $(M c/\hbar \sigma_0)^2
\approx 0.936$. For a given value of $u/c$, Eq.\,(\ref{EEul1})
gives the profile of the solitary wave $\sigma(z)/\sigma_0$.
Figure 1 shows such solutions for $u/c = 0.1$ (bottom curve at
$z=0$), 0.4, 0.7, and 0.9 (top).

\begin{figure}
\begin{center}
\includegraphics[width=8cm,height=5.cm]{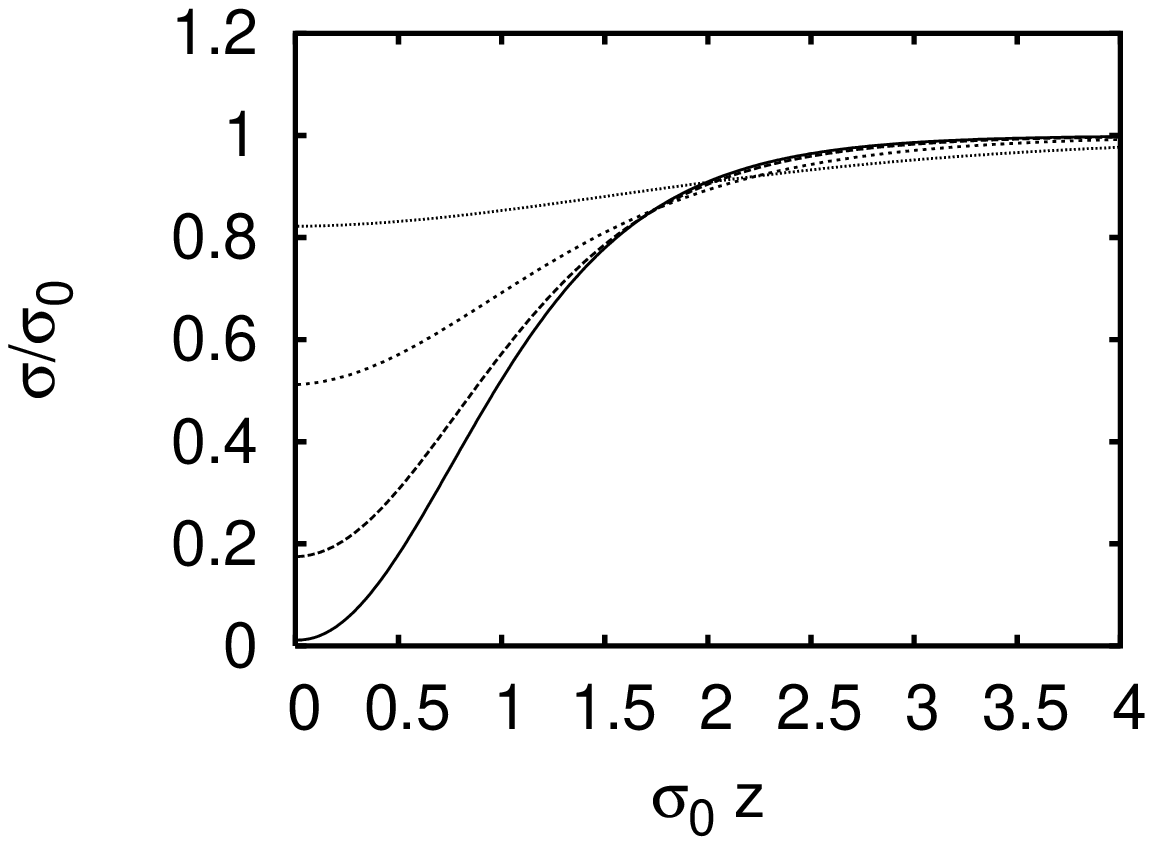}
\begin{caption}
{The density $\sigma(\tilde{z})/\sigma_0$ associated with a
solitary wave given by the solution of Eq.\,(\ref{EEul1}), for
$\gamma_0 = 2 a/(\sigma_0 a_{\perp}^2) = 1$ and $u/c=0.1$
(lowest graph at $z=0$), 0.4, 0.7, and 0.9 (highest), where
$\tilde{z}=\sigma_0 z$.}
\end{caption}
\end{center}
\label{FIG1}
\end{figure}

The energy and the momentum can then be calculated from
Eqs.\,(\ref{een}) and (\ref{pst2a}):
\begin{eqnarray}
{\cal E}(u) = \frac {\hbar^2 \sigma_0^2} M
   \int_{-\infty}^{+\infty}
\left[ \frac {\pi^2 y^3/3 + \kappa y^5}  {1 + 4 y + 4 y^2 +
\kappa y^3} - \frac {\pi^2 /3 + \kappa } {9 + \kappa}
              \right. \nonumber \\
             - 1.471 \, (y - 1) ] \, d \tilde z, \phantom{X}
\label{eeeen}
\end{eqnarray}
and
\begin{eqnarray}
 {\cal P}(u) =
    M u \int_{-\infty}^{\infty}  d\tilde z \, (y-1)^2/y.
\label{pppst2a}
\end{eqnarray}
The dispersion relation ${\cal E}={\cal E}({\cal P})$,
initially obtained by Lieb, is shown as the continuous (lower)
line in Fig.\,2.  The maximum value of $\cal P$ corresponds to
a dark solitary wave with $u=0$.  For long wavelengths, the
solitary waves become ordinary sound waves with the usual
linear dispersion relation \cite{Tsuzuki,KPa,JK,KPa2}.  From
Eq.\,(\ref{ton5}) one can also calculate the dispersion
relation appropriate for the Bogoliubov mode.  This has the
usual form
\begin{eqnarray}
   {\cal E} = \sqrt{\left( {\cal P}^2/ 2M \right)^2 +
   (c {\cal P})^2}.
\label{BBO}
\end{eqnarray}
The Bogoliubov dispersion relation, ${\cal E}={\cal E}({\cal
P})$, is shown as the dashed (higher) line in Fig.\,2.

\begin{figure}
\begin{center}
\includegraphics[width=8cm,height=5.cm]{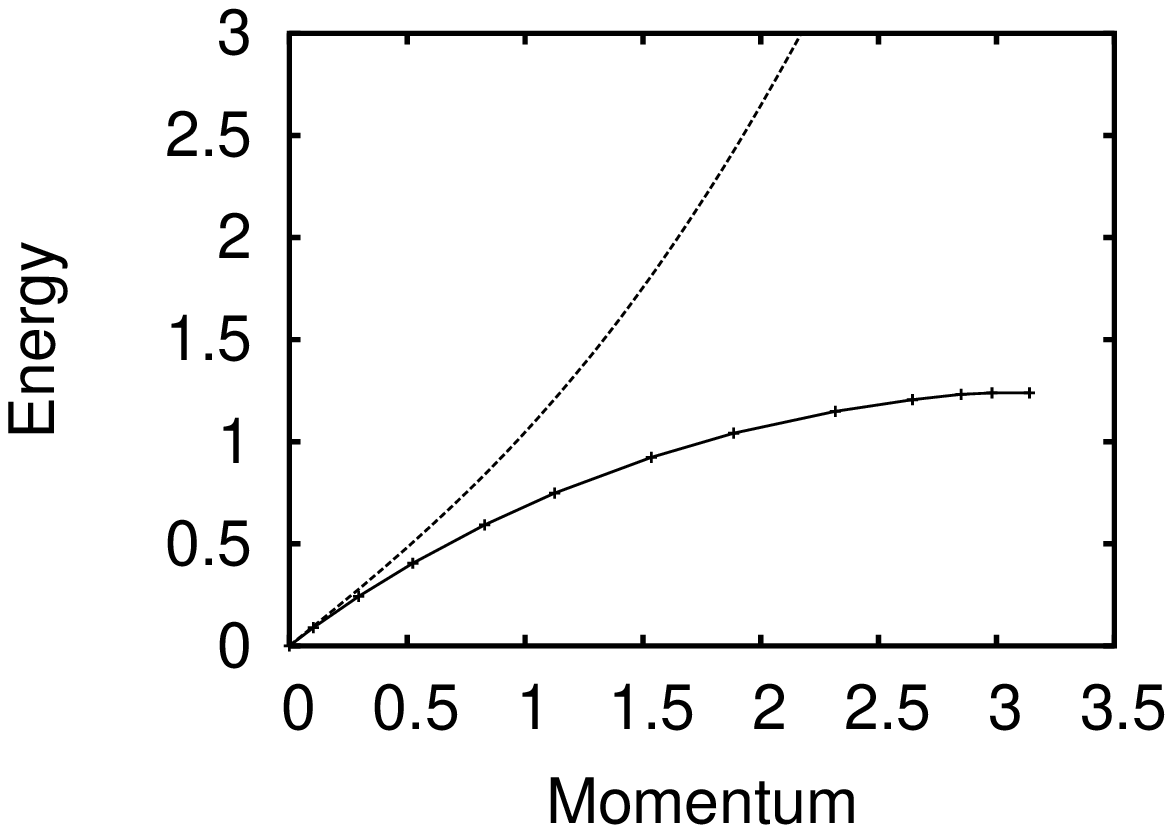}
\begin{caption}
{The dispersion relation ${\cal E} = {\cal E}({\cal P})$ for
solitary waves (lower) and for the Bogoliubov mode (higher). The
momentum is measured in units of $\hbar \sigma_0$, and the energy
is measured in units of $\hbar^2 \sigma_0^2/M$. Here $\gamma_0 = 2
a/(\sigma_0 a_{\perp}^2)=1$.}
\end{caption}
\end{center}
\label{FIG2}
\end{figure}
\begin{figure}
\begin{center}
\includegraphics[width=8cm,height=5.cm]{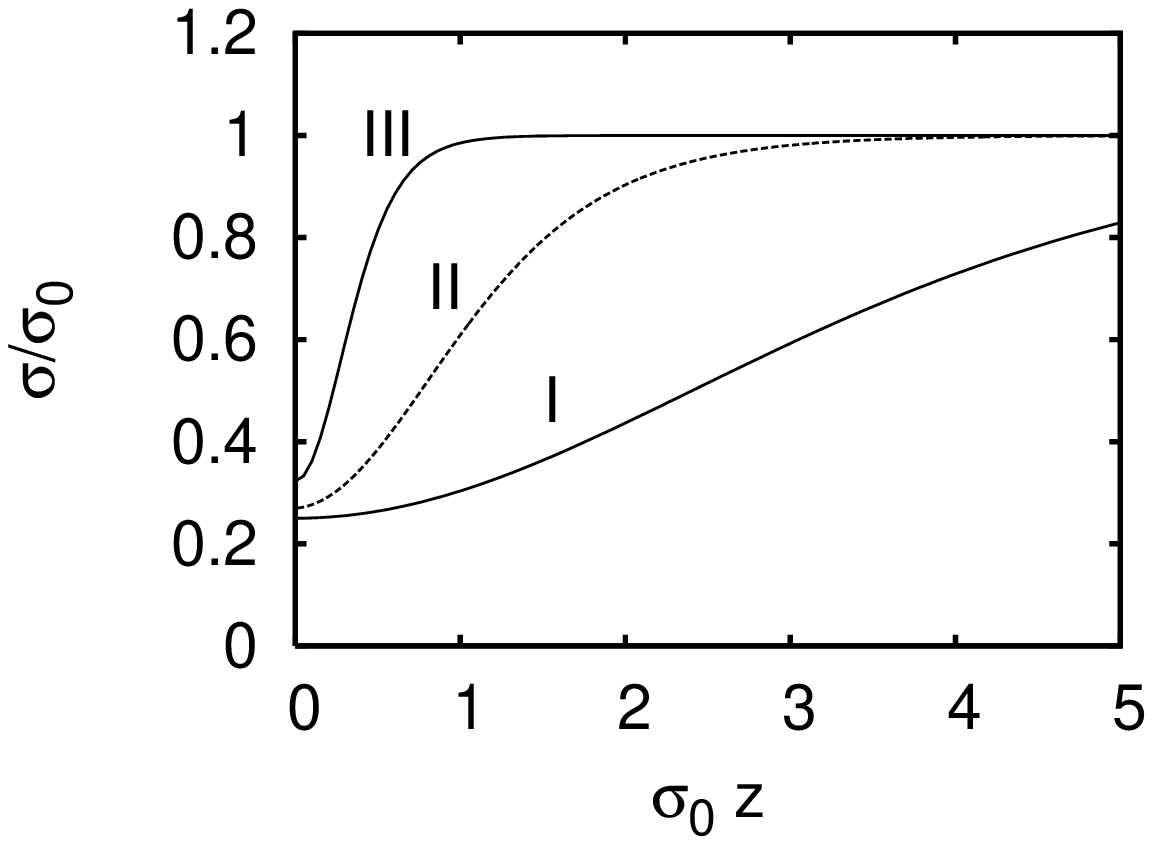}
\begin{caption}
{The profile of the solitary waves for $u/c=1/2$ and three
values of $\gamma_0= 2 a/(\sigma_0 a_{\perp}^2)$. In the lowest
one (I) $\gamma_0 = 1/10$, corresponding to the one-dimensional
Gross-Pitaevskii limit, Eq.\,(\ref{solEull}), in the middle one
(II) $\gamma_0 = 1$, corresponding to the intermediate regime,
Eq.\,(\ref{EEul1}), and in the highest one (III) $\gamma_0 =
10$, corresponding to the Tonks-Girardeau limit,
Eq.\,(\ref{ssolEull}).}
\end{caption}
\end{center}
\label{FIG3}
\end{figure}

The two conditions derived above establishing the typical
values of $\sigma$ that characterize the three regimes,
$\sigma_{c,1} a \sim 1$ and $\sigma_{c,2} a_{\perp}^2 \sim a$,
give values of $\sigma$ which differ by roughly one order of
magnitude when the transverse confinement is strong as in
Ref.\,\cite{Science}. For example, for $a \sim 100$ \AA \, and
$\omega_{\perp} = 2 \pi \times 70.7$ kHz (which implies that
$a_{\perp} \sim 0.04\, \mu$m), we see that $\sigma_{c,1} \sim
10^6$/cm and $\sigma_{c,2} \sim 10^5$/cm.

Another remarkable feature of this problem is that it is
independent of the scattering length in the Tonks-Girardeau
limit.  Specifically, in the two limiting cases (i.e., the
one-dimensional Gross-Pitaevskii equation and the
Tonks-Girardeau equation), the characteristic widths of
solitary waves is on the order of the coherence length $\xi
\sim a_{\perp}/ (\sigma_0 a)^{1/2}$ and on the order of the
interparticle spacing $1/\sigma_0$, respectively.  This allows
us to understand the qualitative features of the profiles shown
in Fig.\,3.  For the largest value of $\sigma_0 = 10 \,
\sigma_c$, which explores the Gross-Pitaevskii limit, the
characteristic width of the disturbance (in units of
$\sigma_0^{-1}$) is on the order of $(\sigma_0
a_{\perp}^2/a)^{1/2} \sim 1/\sqrt{\gamma_0}$, which is larger
than unity.  In the opposite Tonks-Girardeau limit with
$\sigma=\sigma_c/10$, the width is of order unity in units of
$\sigma_0^{-1}$. Therefore, as one moves from one regime to the
other, the width of the solitary wave in the dimensionless
units of Fig.\,3 decreases and eventually reaches a constant
value of order unity.  This fact can be used as an experimental
signature of the transition from one limit to the other.

The present model is valid provided that the solitary waves
have a width which is larger than the atom-atom spacing.  While
our predictions are thus expected to be reliable for waves
whose size is larger than $1/\sigma_0$ (e.g., sound waves),
they are pushed to the limits of their validity in the extreme
cases of the Tonks-Girardeau limit ($\gamma \gg 1$) and for
narrow (and thus slow) pulses.  Predictions in these regions
are expected to be qualitatively, but not quantitatively,
correct \cite{Kol3}. The Boboliubov spectrum, on the other
hand, agrees with the exact result \cite{Lieb} for ${\cal P}
\to 0$ and ${\cal P} \to \infty$ for all values of $\gamma$.

Finally, as seen in Fig.\,2, the dispersion relations for the
two modes converge in the limit of long wavelength since both
correspond to small amplitude sound waves in this limit.  For
shorter wavelengths, however, they diverge.  The characteristic
momentum, ${\cal P}_c$, for which differences appear is on the
order of $\hbar/\xi$. Thus, ${\cal P}_c/ {\cal P}_0 \sim
\sqrt{a/(\sigma_0 a_{\perp}^2)} \sim \sqrt{\gamma_0}$, which is
on the order of unity at the transition and larger than unity
as one approaches the Tonks-Girardeau regime.  This behavior is
quite different from that between the three-dimensional and the
one-dimensional problems studied in Refs.\,\cite{KPa,JK,KPa2},
where the two modes have different energies and momenta.  In
this respect the present problem more closely resembles the
homogeneous problem initially studied by Lieb
\cite{Lieb,Tsuzuki}.

We thank Nikos Papanicolaou and Stephanie Reimann for useful
discussions. M.\"O. thanks Peter Drummond for presenting 
the Lieb-Liniger model to him. We also acknowledge financial 
support from the European Community project ULTRA-1D 
(NMP4-CT-2003-505457).

\end{document}